# Hemoglycin as a Candidate for Permanent Memory

Julie E. M. McGeoch[1] and Malcolm W. McGeoch[2]

[1] High Energy & Optical and Infrared DIVs, Smithsonian Astrophysical Observatory Center for Astrophysics | Harvard & Smithsonian, 60 Garden St., MS 70, MA 02138, USA.
[2] PLEX Corporation, 275 Martine St., Suite 100, Fall River, MA 02723, USA

**Abstract**
The Hemoglycin polymer of glycine derived from space is terminated by iron and silicon to form a three-dimensional regular lattice with a high density of silicon tetrahedral vertices. Divalent cation bridges bypass vertices to facilitate electron conduction axially from layer to layer of the lattice. In principle, the exact distribution of Mg or Ca ions around each vertex can encode a very large amount of information. In space where the polymer forms in a molecular cloud the current derived from photon input will fade but in-fall hemoglycin in life forms on habitable planets like Earth, could be contained behind lipid electronic barriers of the nervous system in neurons or glia. Energy minimization of the permutations for magnesium or calcium occupancy showed deeper binding for calcium-dominated configurations. A low energy state with two independent magnesium to calcium substitutions appears to be key to the maintenance of long-term chemical modification. This state could be read via a pulse train that induced the matching spatial distribution of calcium to render adjacent insulating lattice regions conductive and provide a completed conduction path, signaling recognition. Writing of information may involve spatial modulation of the Na/Ca exchanger within the lipid membrane.

**Introduction**
Permanent memory in the human brain lasts on average eighty years and the mechanism of embedding and recalling permanent information is not yet known. Here is described a candidate molecule termed Hemoglycin, that could be the basis of permanent memory. The actions of calcium and magnesium in memory relate directly to their interactions with hemoglycin. This polymer, first identified in meteoritic material [1,2], can form a three-dimensional lattice comprising 4-way tetrahedral vertices joined by 5nm rod-like hemoglycin molecules [3,4]. The lattice has Diamond-2H structure as defined by its vertex positions in space [4]. Although artificial plane-lattice structures have been devised [5], the 3-dimensional hemoglycin lattice is highly unusual in that it has emerged in nature as an internal component of stromatolite ooids [3], the small calcareous nodules present in these ancient rock-like assemblages. Based upon the recent discovery of hemoglycin in sea-foam [6] it is probable that hemoglycin from cosmic dust has been delivered to Earth continuously over the millenia, leading to the question as to whether its properties could have been relevant to the needs of developing life forms.

In principle, cognition and memory could utilize such a lattice with its open internal passages for access by ionic conduction and its high density of vertices [2] available to be programmed by selective chemical alteration. In the present study we find that indeed the hemoglycin lattice has the potential to perform as a memory module that is controllable by Mg and Ca variations. The same system may also be able to perform as a "hard-wired" processor of visual information.

**Description of the hemoglycin lattice**

The 1638Da rod-like molecule of hemoglycin forms a three-dimensional regular lattice with diamond-2H symmetry of its vertices [2,3,4]. A modified hemoglycin rod with not two, but three oxygens between the pair of iron atoms (Fig. 1) had also been indicated in mass spectrometry ([1]: ibid. Figs. 12A, 13A and Table S7.1). We denote the latter rod by the term "1638 + 2(O)" taking into account that there are two such iron-oxygen motifs, one at each end of the rod. This rod type enables up to three divalent cation bridges between any one rod and the three other rods at the same tetrahedral vertex in the lattice, as shown for example in Fig. 1.

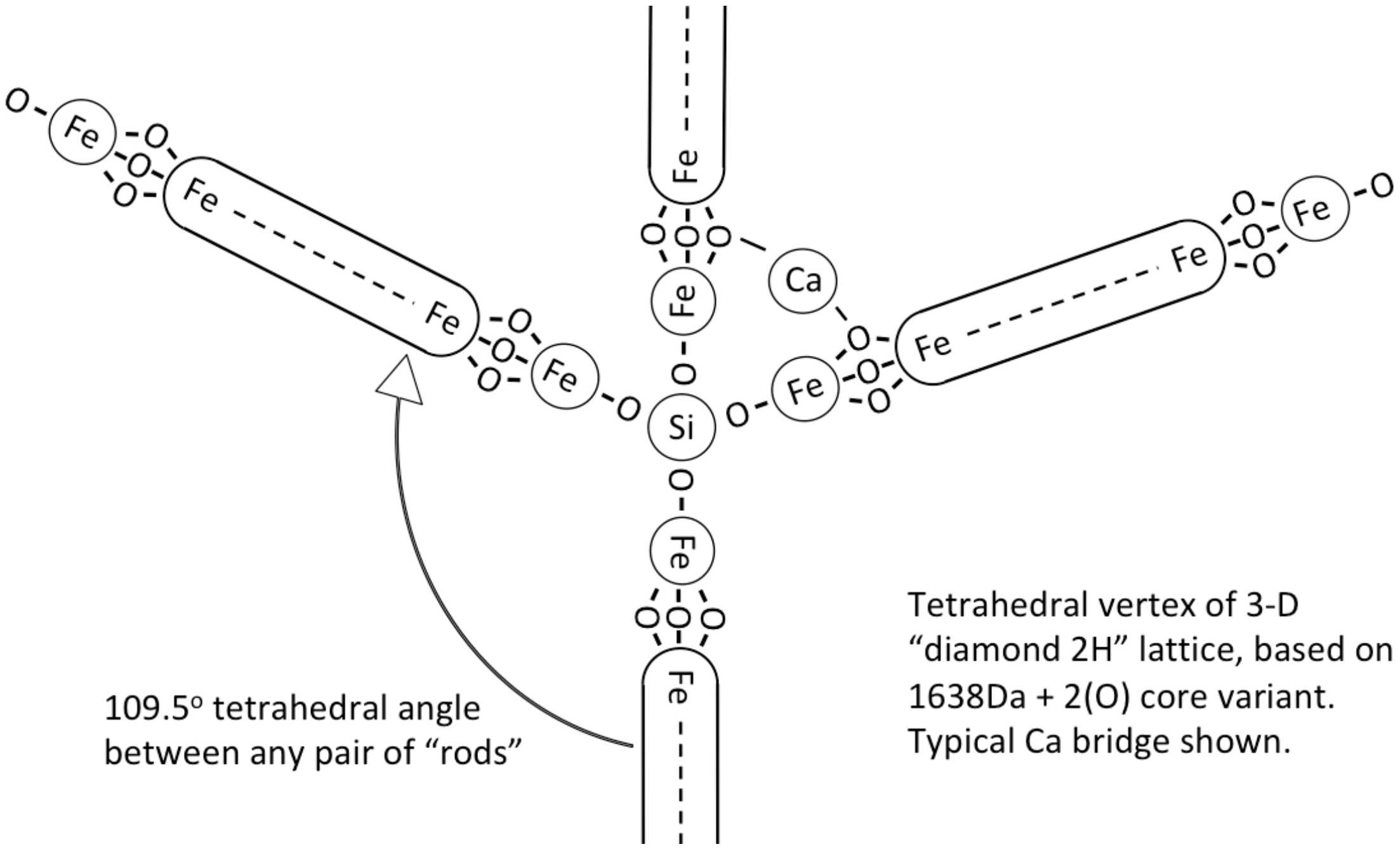

**Fig.1 Vertex of a diamond 2H lattice with four converging 1638+2(O) type rods and a typical divalent cation bridge.**

The lattice structure may be developed from a plane layer of hexagons as shown in Fig. 2 with further details of its geometric properties given in [2,3,4]. Lattice vertices are of either the "Up" type or the "Down" type as illustrated in Fig. 2.

The tetrahedral convergence of four hemoglycin rods of type 1638Da +2(O) creates a total of six opportunities for Ca or Mg to bind between oxygen atoms in neighboring rods, by-passing the vertex and potentially creating electron conduction paths between the rods.

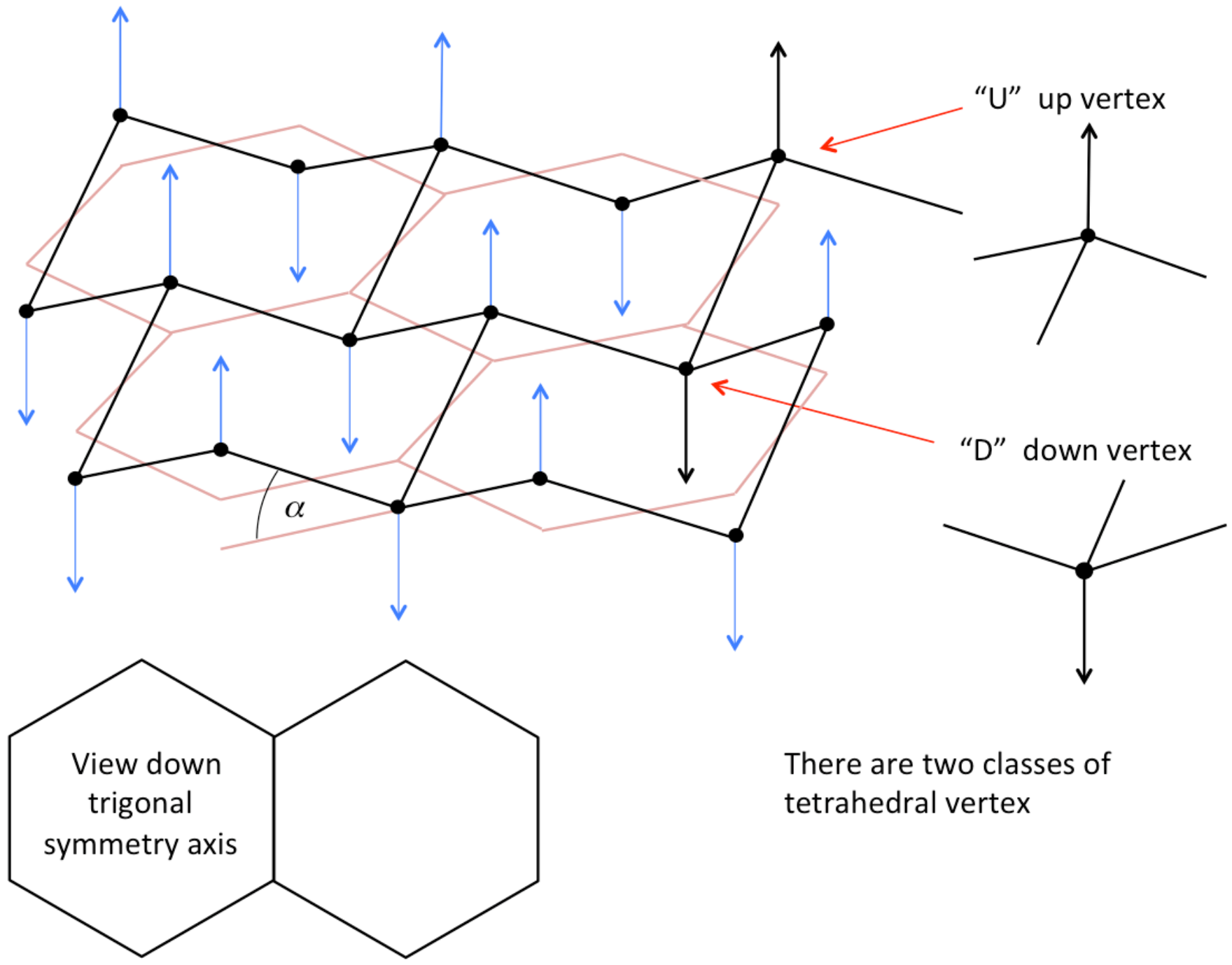

**Fig. 2. Development of a single layer of the lattice starting from a plane of hexagons whose edges are rotated to the tetrahedral inclination illustrating the up (U) and down (D) types of tetrahedral vertex and the trigonal symmetry of the lattice seen axially.**

The present molecular modeling study of hemoglycin was inspired by the apparent utility of its three-dimensional lattice in information storage. The density of its tetrahedral vertices is $5.5 \times 10^{24}$ $m^{-3}$ and they are accessible either by electronic conduction along lattice rods, or ionic diffusion through the seven open through-channels of its structure [2]. Electronic conduction is expected in anti-parallel beta-sheet rods of polymer amino acids [7] but data does not seem to be available for anti-parallel purely glycine chains. Simulation in the Spartan quantum chemical software [8] shows a potential minimum that runs along the line of between-chain hydrogen bonds. This has a rippled floor, but with Fe donating or receiving electrons in the hemoglycin structure there is likely to be electron conduction along the whole length of hemoglycin rods. From a different direction, the anti-parallel beta sheets of spider web exhibit exceptional thermal conduction [9], akin to that of metal, which we take as additional support for electronic conduction in this peptide configuration.

As of writing the full quantum chemistry calculation to confirm bridge conduction is not complete, therefore the balance of this paper is predicated on such conduction being found by simulations, if not by experimentation which would appear to be very difficult. We shall refer to these proposed {oxygen- divalent-cation- oxygen} configurations as "bridges". The six possible divalent bridge locations are illustrated in Fig. 3 where at this point in our discussion either Ca or Mg may be involved in a bridge. Three of the bridges, shown as occupied by Mg, lie approximately in the plane of hexagons while three bridges, shown as occupied by Ca, connect to the axial hemoglycin rod. We adopt the convention of labelling bridges to a rod aligned with the lattice trigonal axis as "axial" bridges, and bridges that lie in a plane perpendicular to the axis, connecting the non-axial rods, as "transverse". These are abbreviated in subscripts to "$_X$" and "$_T$". A detailed view of the atomic composition around one vertex is given in Fig. 4.

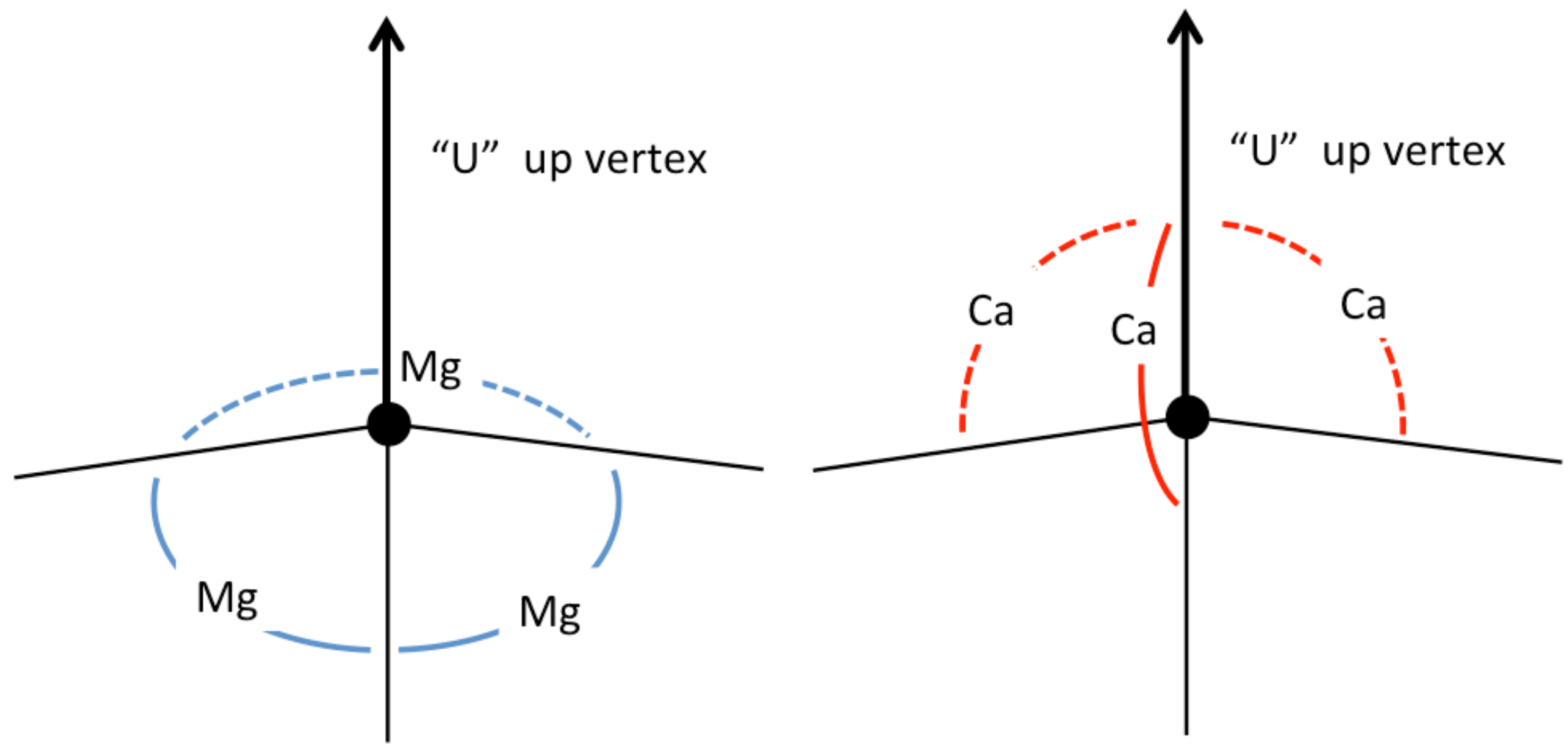

**Fig. 3. Bridge locations around a tetrahedral vertex at the center of four hemoglycin rods. Here, the divalent cations are arbitrarily shown as Mg for the bridges aligned perpendicular to the lattice symmetry axis ("transverse bridges"), and Ca for bridges terminating on rods parallel to the lattice axis ("axial bridges").**

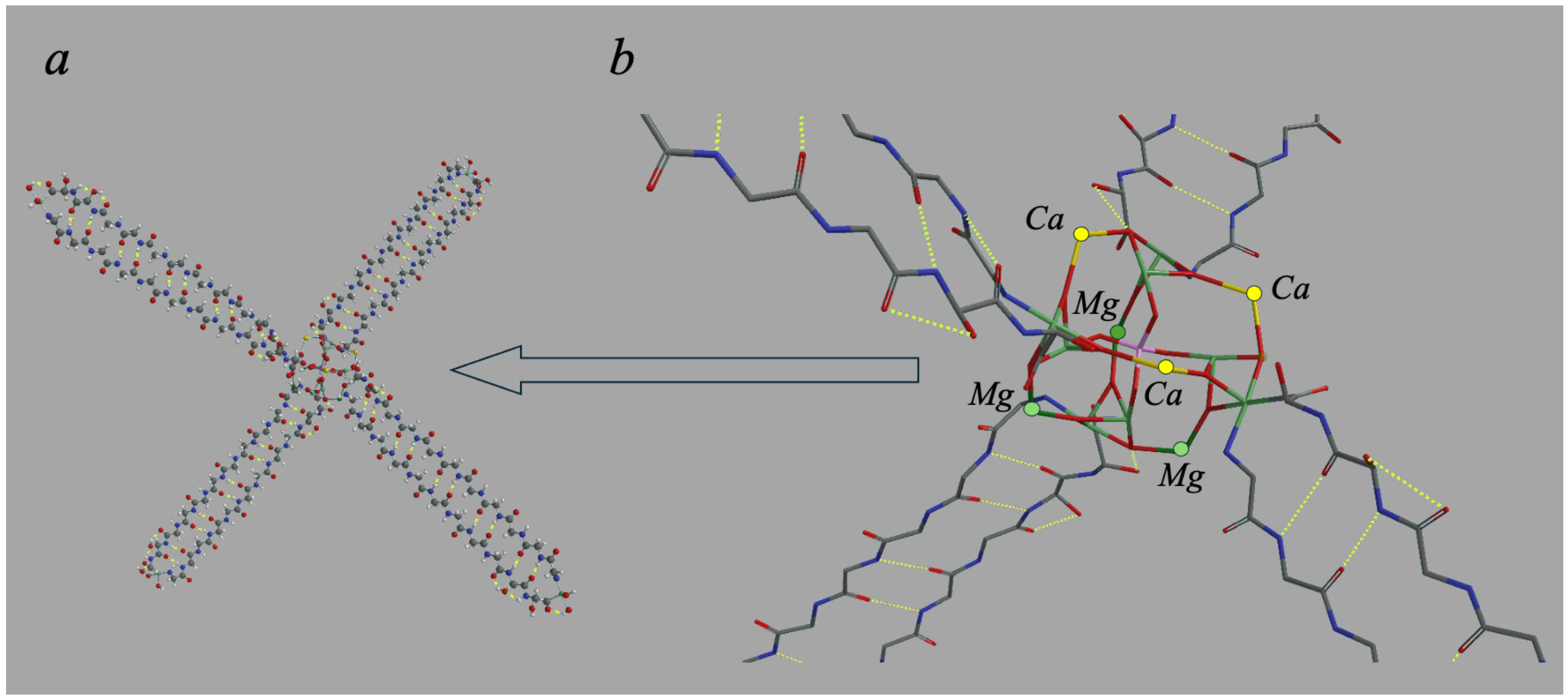

**Fig. 4 (*a*) Spartan rendering of energy-minimized set of four hemoglycin rods around a tetrahedral vertex. (b) Expanded view of the vertex with divalent cation bridges in a state 6 configuration.**

**Axial conduction of electrons through the diamond 2H lattice**

Conduction of electrons may in principle be via divalent cation bridges between the three non-axial rods at a vertex (for example via the Mg ions in Fig 3), but for information processing and storage we believe that electron movement through the lattice will be predominantly axial, assisted by an axial electric field. An axial field has zero driving effect on the three "transverse" bridges, but it does have field components along the three bridges connecting at an inclination to an axial rod, illustrated by the Ca ion bridges (in Fig. 3). From here onward we assume that Ca axial bridges conduct electrons, that is, they do not present a significant potential barrier to an electron wave function connecting two rods.

In Fig. 5 a typical electron is shown advancing from layer to layer of the lattice via divalent bridges. The simulation proceeds in steps, starting below a "D" vertex in a first (M=1) layer, rising through two layers, and finishing above a "U" vertex in layer M=2 (which is below a "D" vertex in layer 3) at which point the lattice repeats (though not necessarily the electron path).

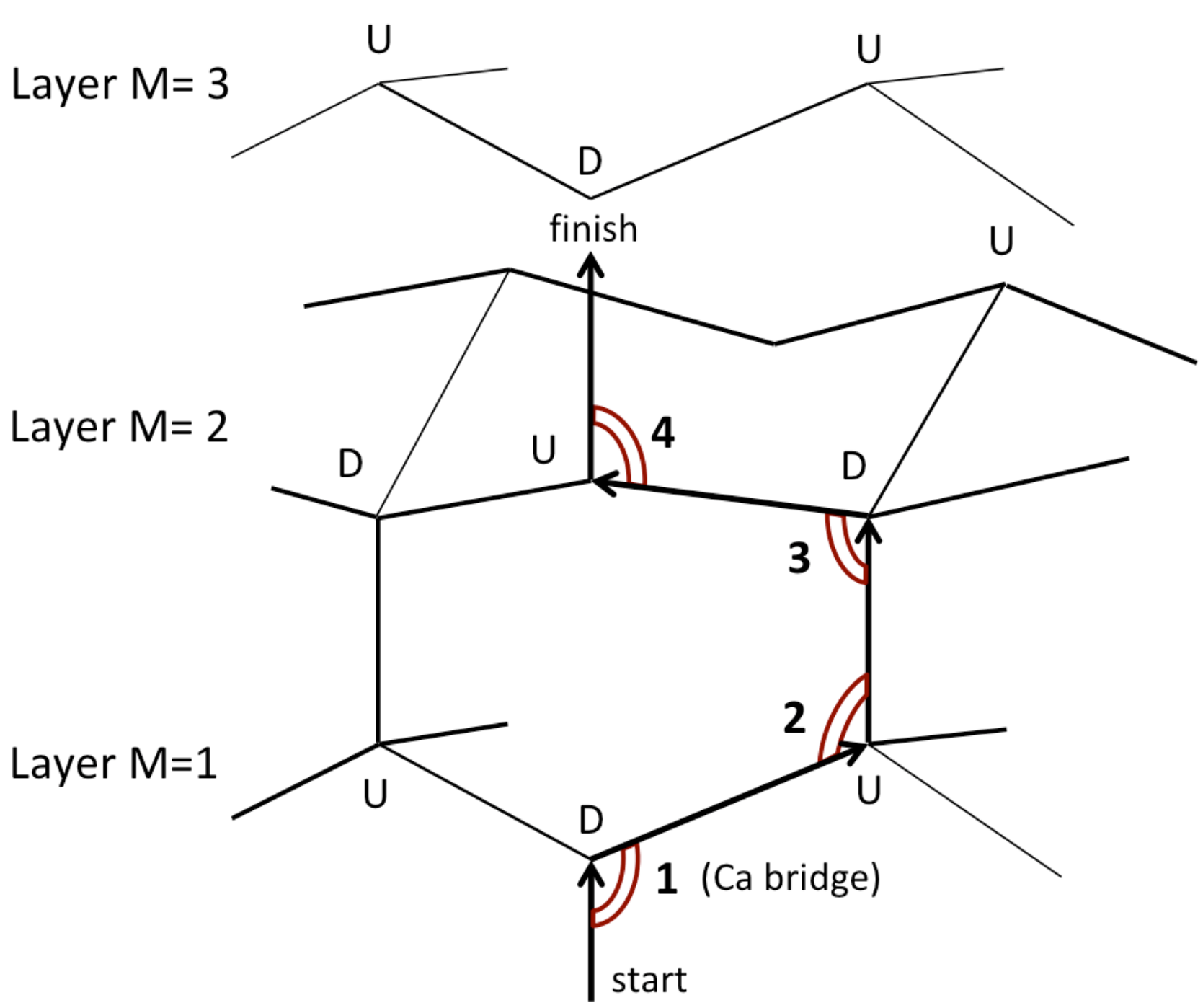

**Fig. 5. Cutaway view of an electron path starting below a "D" vertex in layer 1 and proceeding via axial Ca bridges 1, 2, 3, 4 to just below a "D" vertex in layer 3.**

**Observations concerning the control of electron current via the disposition of calcium bridges at vertices.**

We investigated computationally the passage of electrons axially through a hemoglycin diamond 2H lattice with a varying distribution of Ca bridges connecting transverse to axial rods (Figs. 3 and 4). Initially we considered either one or two axial Ca bridges per vertex, giving each vertex a random assignment of such bridges. This initial choice of either one or two bridges per vertex was influenced by the existence of a 1638Da hemoglycin rod [1,3,6] that had only two oxygen atoms between the terminal pair of iron atoms, as opposed to the 1638Da + 2(O) rods depicted in Fig. 1 which have three oxygen atoms between the two iron atoms.

Electron path simulations ranging between all-single and all-double axial bridge vertices showed that paths terminated within less than about 10 layers. On reflection, this could have been predicted for the double bridge case as follows: With reference to Fig. 5 an electron approaching the D vertex in layer M=1 would have a 2/3 chance of finding a bridge onto one of the three transverse rods connecting to a neighboring U vertex that could allow it to ascend to the M=2 layer. At any of these U vertices there would be a 2/3 chance of landing at an axial bridge, reducing to 4/9 the chance of rising by one layer. Repeated for N layers the $(4/9)^N$ probability of an intact electron path through the whole lattice dimishes very rapidly.
Axial path simulations through a lattice with a probability *p* of three axial bridges per vertex and probability (1-*p*) of two axial bridges, produced conducting paths through high numbers of layers when *p* exceeded approximately 0.8 with a rapid switch between failure and success (Fig. 6). This phenomenon turned out to be fundamental to the process of recognition in the model of memory developed below.

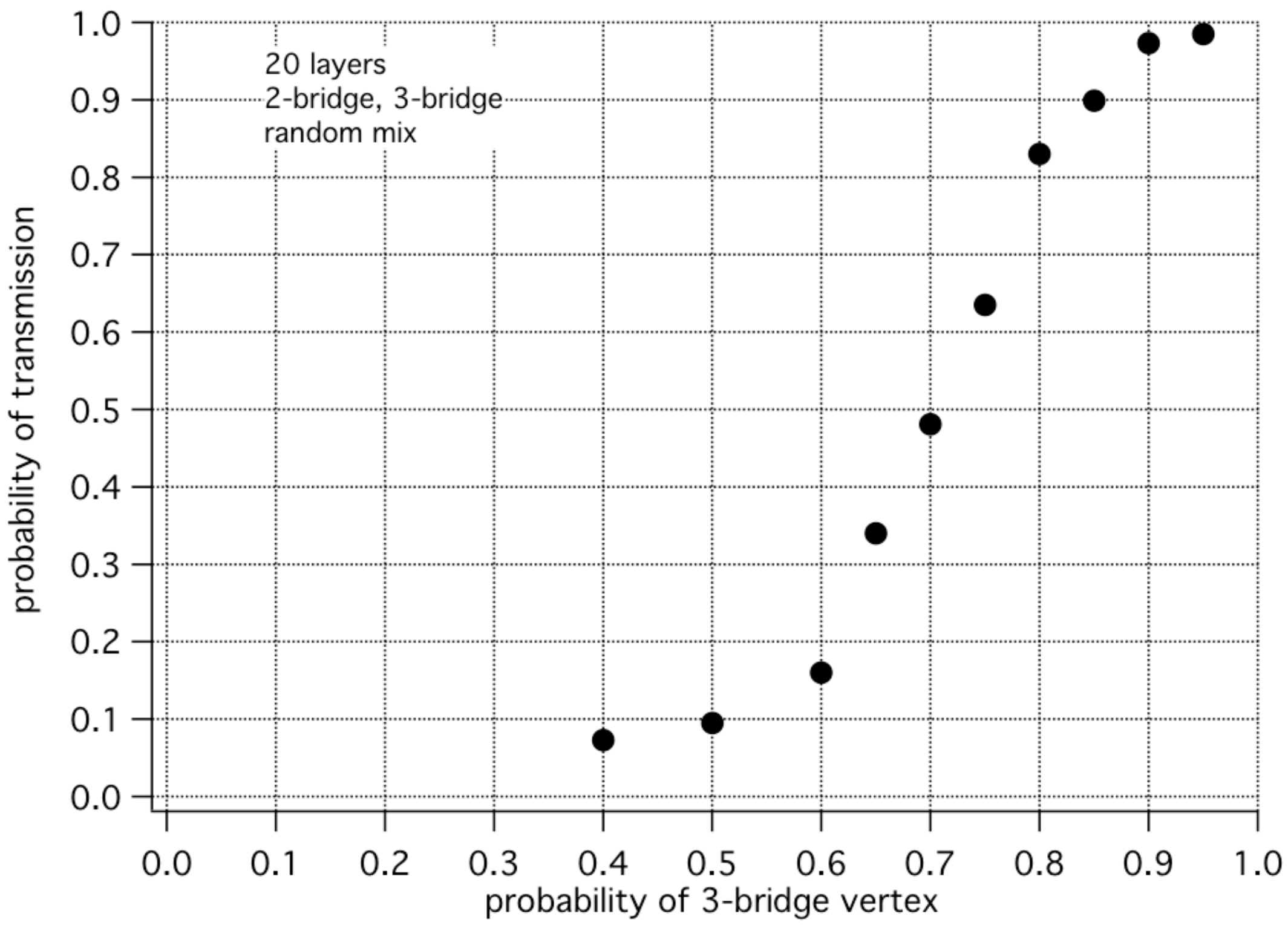

**Fig. 6. Simulation of electron path success through 20 lattice layers as a function of the probability of a vertex having three axial Ca bridges versus two in a random mix.**

**Energetics of Ca and Mg binding to hemoglycin at a vertex**

In order to establish the heirarchy of Ca and Mg binding energies at the six possible divalent cation bridge sites at a vertex, various combinations were evaluated via energy minimization under MMFF [8]. The results, summarized in Table 1, which considers U type vertices (the D type energy set is the same) show generally higher binding strength for Ca than Mg bridges. Surprisingly, the oxygen-oxygen full bridge length was greater for Ca than for Mg in spite of the bonds being stronger, a feature attributable to the larger atomic radius of Ca in this solution-free model environment (the hydrated ionic radii have the reverse dependence). Exploration of different Ca/Mg arrangements yielded a highest binding for the case of $\{(3Ca)_X,(2MgCa)_T\}$ followed by $\{(2CaMg)_X, (2MgCa)_T\}$ where "X" refers to bridges connecting with an axial rod and "T" to transverse bridges connecting laterally within the lattice plane that runs perpendicular to the lattice symmetry axis.

**Table 1 Relative energies of different divalent cation bridge configurations. "x" denotes axial bridges and "T" transverse bridges. All configurations have 6 bridges.**

| Energy kJ/mol relative | State 1 3Mg x 3Ca T | State 2 6Mg | State 3 3Ca x 3Mg T | State 4 6Ca | State 5 2CaMg x 3Mg T | State 6 2CaMg x 2MgCa T | State 7 3Ca x 2MgCa T |
|---|---|---|---|---|---|---|---|
| -700 | | | | | | | |
| -750 | | | | | | | |
| -800 | | | | | | | |
| -850 | | | | | | | |
| -900 | | | | | | | |
| -950 | | | | | | | |

Comparing the all-Ca and all-Mg cases (6 Ca bridges or 6 Mg bridges) there is stronger binding for 6Ca vertices by a total of 59kJ/mol that translates, under the assumption of independence between binding sites (and no binding cooperativity), to an overall difference of approximately 10kJ/mol at an individual site. In the last two columns of Table 1 we obtained substantially deeper binding when one of three transverse Mg bridge atoms was replaced by Ca, possibly indicating cooperative binding. The Table 1 data, generated from a moderately complex model (Fig. 4) via molecular force modeling that did not include solvent interactions, revealed a trend throughout these seven cases toward deeper whole-vertex binding as the ratio of Ca bridges to Mg bridges increased. The energy scale in Table 1 is relative to an arbitrary zero that is a function of the sequence of model assembly in the Spartan software, from the model's individual parts, the largest of which were previously optimized complete hemoglycin subunit cores of mass 1638Da. A least-squares routine was written to extract the most probable Ca and Mg binding energies and remove the arbitrary formation energy, to yield binding energies of 17kJ/mol and 31kJ/mol for Mg and Ca, respectively, with large probable errors. Columns 5, 6 and 7 of Table 1 individually displayed a range of results depending upon the specific bridge ion dispositions, accordingly data from the many permutations was averaged then incorporated into the analysis. The depicted range of binding energy for "state 6" in column 6 develops when Mg is positioned in turn at each of the three axial bridge locations and the lone transverse bridge Ca atom is itself rotated, totaling 9 runs. Likewise in "state 7" in column 7, with completely Ca-filled axial locations, the data spread develops when the lone transverse Ca is rotated through its three possible locations relative to the axial set.

**Memory: storage and recall.**

Here, in view of the energy dependencies in Table 1, we note that the relative Ca and Mg concentrations will tip the balance between the various states, with more significant effect in the more stably bound states in columns 6 and 7 that are separated by energy gaps of up to a relatively large 40kJ/mol in energy. We therefore introduce the concept of physiological control of the lattice via ionic Ca and Mg concentrations, by proposing a topology with a length of lattice axially oriented inside a cylindrical bilayer-lipid-walled tube equipped with pumps, ion channels and exchangers (Fig. 7).

**Hemoglycin in physiological surroundings**

The intracellular Mg:Ca ratio in resting mammalian glial cells lies roughly between 2,000 and 5,000. Without hemoglycin the balance of Ca and Mg in the cytosol within a cylinder of lipid bilayer membrane is set by the action of several pumps and ion channels, the ones most relevant to dynamic Mg and Ca control being:

a) the Na/K ATPase pump that employs the chemical energy of ATP to remove 3Na ions from the cytosol and admit 2K ions. This is the prime mover that maintains a negative internal potential of -70mV relative to the extracellular medium. With a bilayer capacitance of $1\mu F/cm^2$ the membrane stores an energy of $25\mu J/m^2$ available to power events;

b) The Na stimulus channel [10] that admits into the cytosol a burst of extracellular ions via a high conductance non-selective central pore at a hyperpolarizing voltage. The pore is rapidly closed by the cooperative binding of Ca rising at the cytosolic side [10,11];

c) The Na/Ca exchanger [12,13] that carries one doubly-charged Ca ion out when powered by the inward movement of 3Na ions. This operates more rapidly when the driving trans-membrane voltage difference increases.

Other channels are present, notably the K-permeable channels that contribute to the voltage equilibrium without directly consuming energy.

The stimulus channel of the mammalian brain has been observed in patch clamp measurements to generate oscillation between 0.2 and 700Hz [10], and will be assumed here to be stationed within the lipid wall of a memory-encoding cylinder, responding to an incoming hyperpolarizing voltage pulse train and admitting Ca in a spatial pattern along the length of the hemoglycin cylinder, to replace Mg occupancy in selected divalent bridge sites on hemoglycin.

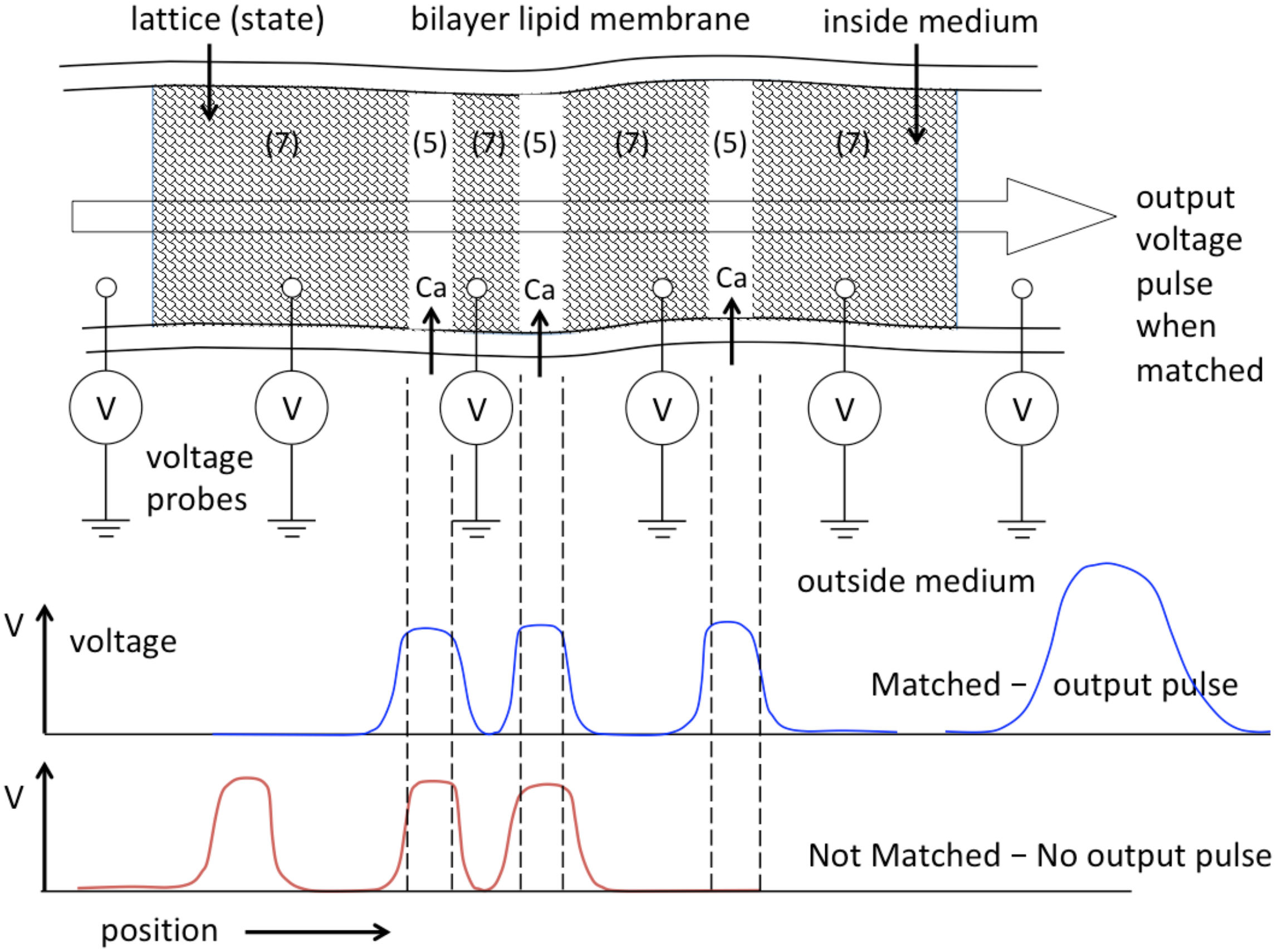

**Fig. 7 A cylinder of lattice is axially oriented inside an insulating lipid bilayer membrane. Regions in state 5 are a previously imprinted "memory". The exact matching voltage pattern, incident later, triggers hyperpolarization Na/Ca channels that raise cytosolic Ca and locally convert state 5, via state 6 to state 7 to create a conducting path through the length of lattice, emitting a voltage pulse.**

The occupancy of a hemoglycin divalent bridge site, whether by Mg or Ca, is determined by their relative ionic concentrations, relative ionic binding energies to hemoglycin, and the temperature, which we will assume to be the physiological 310K.
Individual divalent bridge configurations follow an equilibrium approximately governed by:

$$\frac{[L]}{[M]} = \frac{[Mg]}{[Ca]}\exp\left(-\frac{E_{LM}}{kT_P}\right) \qquad (1)$$

in which divalent configuration (state) ($L$) has more Mg than configuration ($M$), $E_{LM}$ is the binding energy difference taken as positive for deeper Ca binding than Mg, and $kT_P$ is the Boltzmann factor at the physiological temperature of 310K ($kT_P$ = 2.58kJ/mol).

From (1), setting $[L] = [M]$ to denote the point of transition from Mg to Ca occupancy, with [Mg]/[Ca] = 5,000 we find $E_{LM}$ needs to exceed 22kJ/mol. This lies within the general range of the Ca/Mg substitution energy differences in Table 1, indicating the possibility of state transitions between elements of states 5, 6 and 7 at close to the physiological Mg/Ca ratio in spite of considerable energy differences.

**Scope of information storage**

At a vertex there are two possible occupants for each bridge site (Ca or Mg) therefore there are $2^6$ in-principle combinations when all bridges are occupied. These break down into two classes "axial" (X) and "transverse" (T), 3 bridges in each. If all are occupied, there are four distinguishable combinations (3Ca, 2CaMg, Ca2Mg and 3Mg) within each class. In the whole vertex, with 4 in each of class X or T there are 16 (X+T) combinations ranging through ($3Ca_X3Ca_T$ …. $3Mg_X3Mg_T$).

Lattice axial electron conductivity depends upon a strong >80% dominance of vertices containing $3Ca_X$ plus other "T" groups, over all other classes (discussed above). Transitions between these states obey a sum rule. Naming [J] as the concentration of any of the 16 vertex "types" or "states" there is a sum rule [V]=Σ[J] for all J, relating the concentration of states to the fixed total vertex concentration [V]. Transitions out of one state cause a one-for-one increase in the collective divalent concentrations of the other target states. How fast can these interchanges occur?

**Diffusion and interchange rates**

The large ratio of cytosolic Mg to Ca dictates the relative rates for interchange, which occurs in two steps: a) departure of one divalent cation from its site between oxygen atoms on hemoglycin, and b) arrival of the other divalent cation via diffusion, followed by its binding to the site. The first step has a rate given approximately by:

$$R_{DEP} \approx \nu_{OSC} \exp\left(\frac{-E_B}{kT_P}\right) \qquad (2)$$

where $\nu_{OSC}$ is the oscillation frequency of the divalent cation in its site and $E_B$ its binding energy. Taking an oscillation frequency of $10^{13}$ $s^{-1}$ together with a bridge binding energy of 31kJ/mol for Ca (above) we find a departure rate for Ca of $6x10^7$ $s^{-1}$. For Mg with binding of 17kJ/Mol (above) we would have a departure rate of $1.4x10^{10}$ $s^{-1}$. The other part of an interchange relates to the arrival rate. As the Ca and Mg diffusion coefficients in $H_2O$ are similar, at $7x10^{-10}$ $m^2/s$, their arrival rates are in the ratio of their concentrations. The actual arrival rate depends upon ionic diffusion and is inversely dependent on the square of the relevant diffusion length. At a length of 1µm the characteristic diffusion time is 1.5ms. At 10µm the diffusion rate would be lower and its time would be longer, at 150ms. The hemoglycin transverse dimension could be of the order of 1µm, allowing transverse ionic interchange of Ca or Mg on the 1ms timescale, but there would exist diffusion-limited longitudinal interchange from more distant ionic reservoirs axially spaced by more than 1µm. This time asymmetry appears to aid the "read" and "write" processes discussed below.

We have seen above (Fig.6) that lattice axial electron conductivity depends upon a strong >80% dominance of vertices containing $3Ca_X$ over all other vertex states. "Recall" is proposed to occur via a switch from insulating to conducting of a length of lattice that is enclosed by a wall of lipid bi-layer, the switch occurring when simultaneously along the whole lattice length [$3Ca_X$] exceeds about 0.8[V] where [ ] denotes a concentration of vertex states.

**“Read” and “Write” mechanisms**
First we propose a “read” mechanism that is reached by simple considerations. With reference to Fig. 7, prior to the arrival of a train of voltage pulses (the signal), an imprinted “memory” comprises a set of short lattice regions distributed axially that are lacking in sufficient $3Ca_X$ vertices for through electron passage, i.e. a lattice with $[3Ca_X] < 0.8$ [V] in parts of its length.

It is proposed that as the signal progresses along the membrane and lattice it opens hyperpolarization Na (with Ca) stimulus channels that admit Ca briefly wherever the voltage is high before being closed by the cooperative binding of Ca on the cytosolic side. The signal creates a moving internal set of Ca impulses that locally increase the concentration of vertices with $3Ca_X$. If the signal spatially matches the imprinted pattern of vertices lacking in $3Ca_X$, the non-conducting parts of the lattice are rendered conducting, and a large current pulse, carried by electrons, can pass along the lattice, an outcome perceived as an exact superposition of the signal on a prior imprinted pattern.

With present knowledge about the divalent binding energy heirarchy (Table 1) there is not sufficiently precise knowledge of $E_{LM}$ factors (Equ. 1) to be certain which states are most stable, yet sufficiently malleable, to contain the imprinted pattern for long periods at the resting cytosolic Mg:Ca ratio. State 5 may be of interest.

It is noted that the information contained at a vertex in state 5 is doubly encoded. Two interchanges are required to reach $3Ca_X$ state 7. The 5 -> 6 transition involves repacement of a transverse Mg by Ca, and the 6 -> 7 transition involves replacement of an axial Mg by Ca (Table 1). Reading via a temporary transition from state 5, via state 6 to state 7, changes Mg into Ca twice. When the Ca pulse subsides via Na/Ca exchanger action and state 7 reverts to states 6 and 5 the “memory”, inscribed in Mg, is almost unchanged.

**“Write” mechanism**
Our proposed “write” mechanism is more tentative than the read mechanism already described. Prior to “writing” one has to consider a spatially uniform lipid bilayer cylinder containing an axially aligned hemoglycin lattice, but one with a spatially uniform distribution of vertex states in accord with the resting cytosolic Mg/Ca ratio. This would be our “blank sheet”. It would have slightly under 80% 3CaX vertices and therefore be non-conducting to electrons.
As before, a pulse train of hyperpolarizing voltages (encoded sound or visual information) would enter the bilayer tube. The corresponding pattern of Ca impulses would be delivered in a moving procession, raising all the internal vertices to the conducting state 7 by the time the procession had reached the end of the bilayer cylinder. At this time there would be a spatially differentiated set of stimulus channels delivering Ca. When the hemoglycin lattice (dominated now by state 7) presented its short circuit via electron conduction, the driving force for ionic conduction would vanish, immobilizing the Ca entry positions and allowing Ca entry to continue for a short period only where it had been initiated.
Two things could follow:

a) The cooperative Ca binding to close the stimulus channel would operate first at the open pores where most Ca had been delivered.
b) If the Na/Ca exchanger were to increase its density in the membrane locally around pores where highest Ca had been experienced, then a slower form of imprinting could take place, so that subsequent arrival of the same voltage train would not be able to alter the cytosolic Ca to the same extent where a pore delivered it. Elsewhere along the axis, the relatively lower density of exchangers would result in an increased resting cytosolic Ca, tipping the vertex states in these other regions more toward state 7, at the expense of states 6 and 5. This would prime the hemoglycin lattice for the read process described above.

Although the voltage pulses would have their effect in milliseconds, the timescale for an exchanger density increase in the membrane would be much longer, possibly seconds and upward, depending upon whether existing exchangers needed to be moved within the membrane, or new ones synthesized and transported there.

In summary, the "write" mechanism could involve concentration of Na/Ca exchangers locally in the membrane in an axial density pattern that matched the incoming voltage pulse train signal, leaving a deficit of Ca and therefore of state 7 in just those locations when the cytosol was at rest, providing the optimized situation for our proposed calcium injection "read" mechanism above.

**Discussion**

In the neuroscience field of permanent memory in a healthy brain no candidate molecule for permanency has been presented and we, working on the structure of space molecule hemoglycin, are now able to describe its potential information storage role that is based on the three-dimensional lattice structure that it forms. Without a model for its function the neuroscience community has heavily associated permanent memory with diseases where memory fails such as Alzheimer's. Current understanding of the amyloid pathology, reviewed by Ono and Tsuji [14], reveals clearly that there is membrane degradation in terms of its fluidity and capability for insulation, accompanied by loss of calcium control, due to the insertion of pores composed of amyloid. All of these findings relating to membrane disruption are consistent with interference in the mechanism for permanent memory proposed in the present paper.

Divalent cations were proposed in order to provide electron conducting bridges from rod to rod and potentially provide by their presence or absence the means to encode pathways through the lattice. An association of Ca with the hemoglycin lattice had been revealed in X-ray scattering from stromatolite ooids [3], suggesting Ca as a candidate ion for the bridges. Because of its high abundance in space relative to Ca, we also included Mg as a candidate bridging ion. Molecular cloud Mg to Ca ratios rise to 100:1 or greater.

To date, the degree of conductivity of these proposed Ca or Mg bridges has not been measured, and is expected to be a challenging quantum chemistry problem on account of the large grouping of atoms in and around a vertex that includes eight Fe atoms. The present work has proceeded on the assumption that because of its deeper binding energy an axial bridge occupied by Ca has at most a small energy barrier to impede electron flow from rod to rod, otherwise electron motion could be slowed by hopping over a series of small energy barriers, but not necessarily reduced to unimportant levels.

Upon introduction of a full set of six Ca bridges the MMFF energy minimization achieved improved tetrahedral orientation of the four rods at a vertex, the free outer ends of the rods becoming almost equidistant from each other to within an error of 0.5nm relative to their length of 5nm. This tendency to alignment suggests that diamond 2H self-assembled lattice formation in stromatolites could have happened swiftly after infall of hemoglycin into the sea, aided by Ca ions [3].

It also was assumed that electron motion would occur in a dominantly axial direction through the lattice as an axial electric field would have driving components parallel to "axial" divalent bridges, but zero component parallel to "transverse" bridges. We followed via a Monte Carlo simulation the conduction paths through lattices with a varying proportion of vertices randomly having either two axial Ca bridges or three axial Ca bridges and found a sudden switch from insulating to conducting when the proportion of three bridge vertices exceeded 80%, a finding that provided a means to retrieve information stored in two-Ca bridge state 5 or 6 via a calcium impulse to create three-bridge state 7 and hence conduction in a stored pattern of previously blocked spatial zones of the lattice.

Many echoes appear here in relation to present day theories of the function of memory, in particular the appearance of Ca in the recall process and the necessary Mg content of the brain [15]. A lattice with states 5, 6 and 7 will display "plasticity" as signals are received, re-inforced, and recognized at a later time. The details will depend upon the relative and absolute Ca and Mg concentrations.

There is a significant Fe content with 8 atoms per lattice vertex. This may have a connection with the Fe content of amyloid plaques [16] that varies from $Fe^0$ to $Fe^{3+}$ in sampled regions. The Fe oxidation state in 1638 + 2(O) vertices is expected to be 2 or 3.

The possible need for patterning of the Na/Ca exchanger in the "write" process (above) may be relevant to the extensive correlations between exchanger type and function in neurological diseases including Alzheimer's [17].

It is possible that the same cellular entity containing hemoglycin could perform calculations with longer pulse trains that encoded a significant load of information. Most obviously, coincidence detection could be performed. For a lattice cylinder of diameter 1μm and length 100μm the time and distance limits imposed by diffusion over lengths longer than about 1μm (discussed above) would enable pulse trains with up to 100 voltage spikes to be memorized or used in a computation.

Whether life has made use of hemoglycin in memory remains to be determined. Hemoglycin has its origin in the pre-solar molecular cloud and from its extraterrestrial isotopes in stromatolite and sea foam does not appear to form on Earth, but rather would have to be acquired by organisms through ingestion after its in-fall. The presence of elevated $^{15}N$ would be a permanent indicator of hemoglycin in brain tissue, as equilibration with terrestrial H would have removed the extra-terrestrial D signature in samples. Mass spectrometry of brain tissue from memory-intensive regions would reveal typical m/z fragments according to established patterns [1,6].

The present theoretical excursion will hopefully inspire further work on that brilliant aid we all have: long term memory.

DATA AVAILABILITY: The code that supports the findings of this study is openly available in Harvard Dataverse at "Hemoglycin as a Candidate for Permanent Memory"and is available from the corresponding author upon reasonable request.

## S Section

### S1 Trigonal Axes

The diamond 2H lattice may be described in terms of x,y,z coordinates, which is useful for calculation of inter-vertex distances in an X-ray scattering analysis [3]. However, a more suitable coordinate system for description and computation of the electron paths as a function of divalent bridge locations is one based on three in-plane axes tilted at 60° to each other, plus one axial numeral to represent layers of the structure. This is illustrated in Figure S1, which represents a first (M=1) layer (equivalent to one layer of distorted hexagons and perpendicular to the trigonal axis of symmetry).

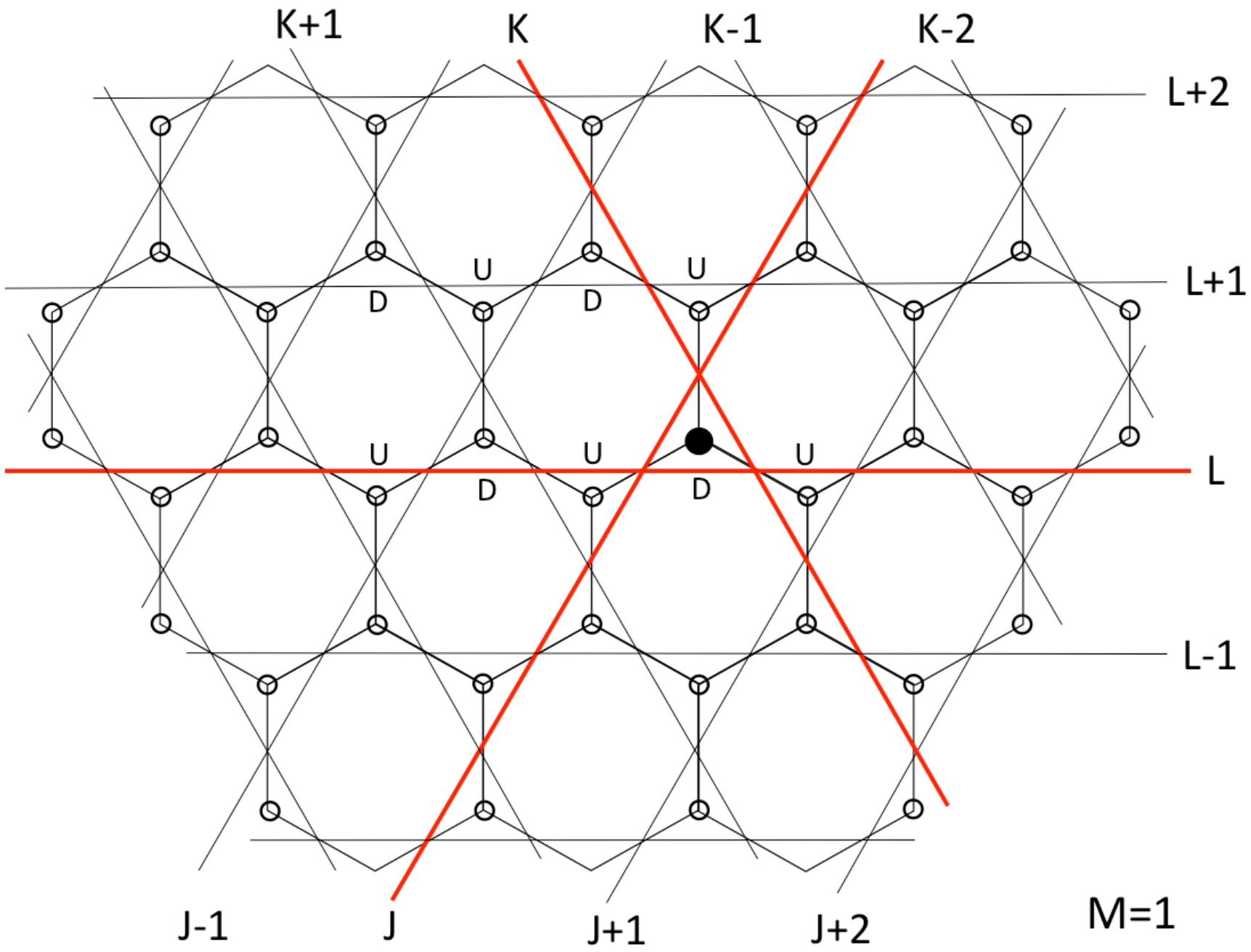

**Fig. S2. A trigonal coordinate system for electron path computation**

For illustrative purposes a "down" (D) vertex is marked. It is seen to lie within a small triangle defined by the J,K,L set of coordinates. All other vertices have their own unique set of J,K,L coordinates, which, together with the layer numeral M, describe the vertex location within a lattice. An electron starting below vertex (J,K,L) in layer M=1 and rising to the D vertex has the chance to travel to three surrounding vertices V(J+1,K,L), V(J,K+1,L) or V(J,K,L+1). This first branching depends upon the disposition of divalent bridges at V(J,K,L). Whether the electron has a bridge at its destination vertex will decide whether it has a path upward above that U type vertex, ending up below a D vertex in layer M=2. Similarly it may progress upward through layer M=2 if bridges exist, to finish below a D vertex in layer M=3. The diamond 2H lattice repeats after two layers, hence the origin of "2" in diamond 2H. The pattern of bridges will vary from vertex to vertex, in the general case.

**S2. Description of the voltage-dependent Na/Ca stimulus channel**
The high-voltage gated sodium/calcium stimulus channel is present in membrane preparations from mammalian brain [10]. It consists of (an estimated) 6 trans-membrane subunits that are identical with the highly conserved subunit C of the ATP synthase also present in mitochondria. It has a large conductance without ionic selectivity (between Na and Ca) and spontaneously oscillates at frequencies up to at least 700Hz. Following its delivery of a small burst of Ca into the cytosol it is shut by the cooperative action of six or more calcium ions binding on the cytosolic side, as pictured in Fig. S2.1 [McGeoch and Palmer (1999)].

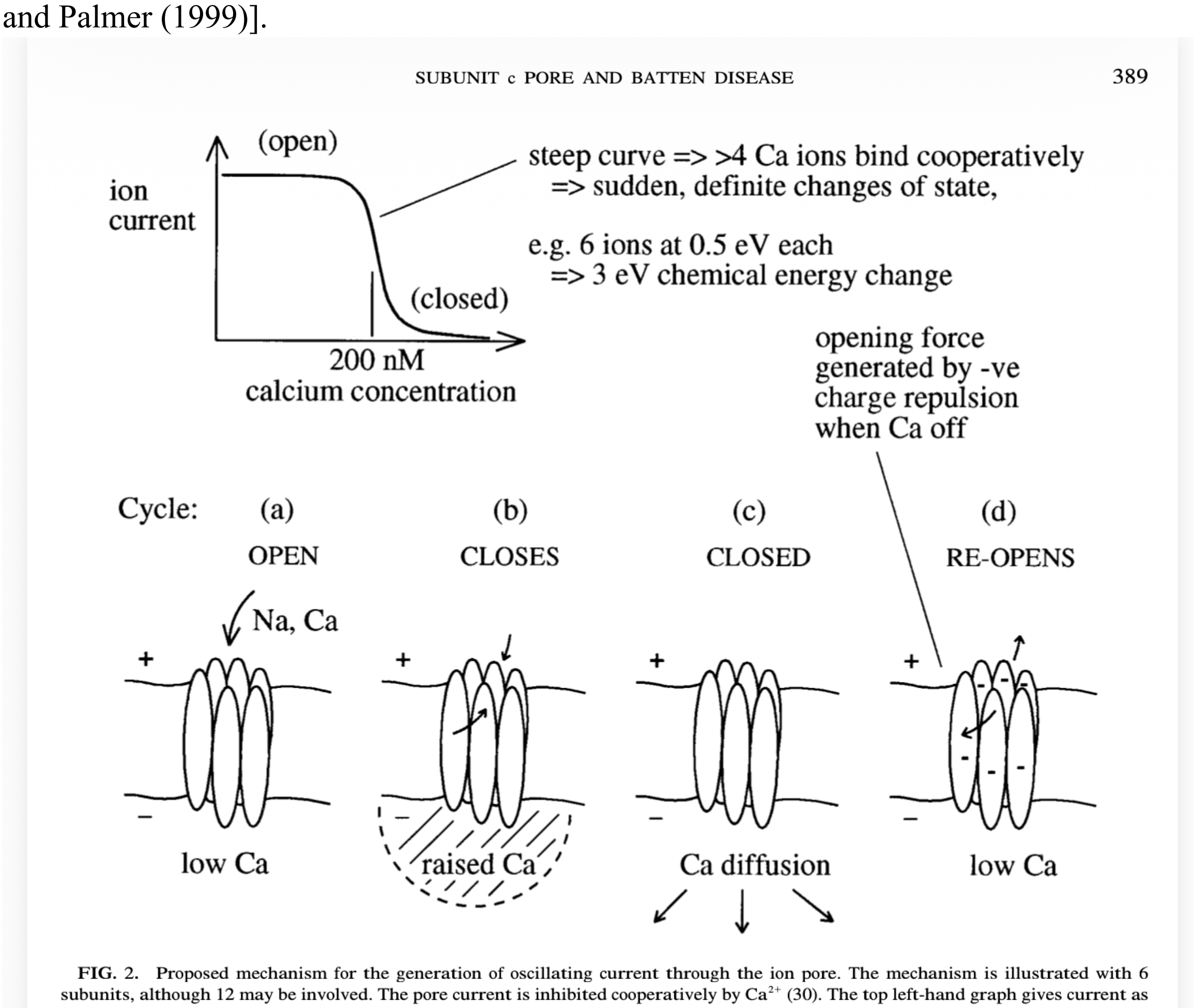
SUBUNIT c PORE AND BATTEN DISEASE 389

**FIG. 2.** Proposed mechanism for the generation of oscillating current through the ion pore. The mechanism is illustrated with 6 subunits, although 12 may be involved. The pore current is inhibited cooperatively by $Ca^{2+}$ (30). The top left-hand graph gives current as a function of calcium concentration showing the steep closing $K_i$ in the range of 200 nM. The energy change with a change of state is given to the right of the graph, and below a schematic diagram shows the role of calcium throughout the oscillation cycle.

Figure from McGeoch and Palmer 1999